\documentclass[journal=jctcce,manuscript=article]{achemso}

\usepackage{chemformula} 
\usepackage[T1]{fontenc} 
\usepackage{siunitx}
\usepackage[version=3]{mhchem}
\usepackage[flushleft]{threeparttable}
\usepackage{adjustbox}
\usepackage{physics}
\usepackage{booktabs}

\author{Christian S. Ahart}
\affiliation[Westlake University]
{Department of Physics, School of Science and Research Center for Industries of the Future, Westlake University, Hangzhou, 310030, China}
\email{chris_ahart@westlake.edu.cn}
\author{Denan Li}
\affiliation[Westlake University]
{Department of Physics, School of Science and Research Center for Industries of the Future, Westlake University, Hangzhou, 310030, China}
\author{Jochen Blumberger}
\affiliation[University College London]
{Department of Physics and Astronomy and Thomas Young Centre, University College London, London WC1E 6BT, UK}
\author{Shi Liu}
\affiliation[Westlake University]
{Department of Physics, School of Science and Research Center for Industries of the Future, Westlake University, Hangzhou, 310030, China}
\email{liushi@westlake.edu.cn}

\title[An \textsf{achemso} demo]
 {Polaron Transport in TiO$_{2}$ from Machine Learning Molecular Dynamics}

\abbreviations{IR,NMR,UV}

\begin{document}

\begin{tocentry}
  \includegraphics[width=1\columnwidth]{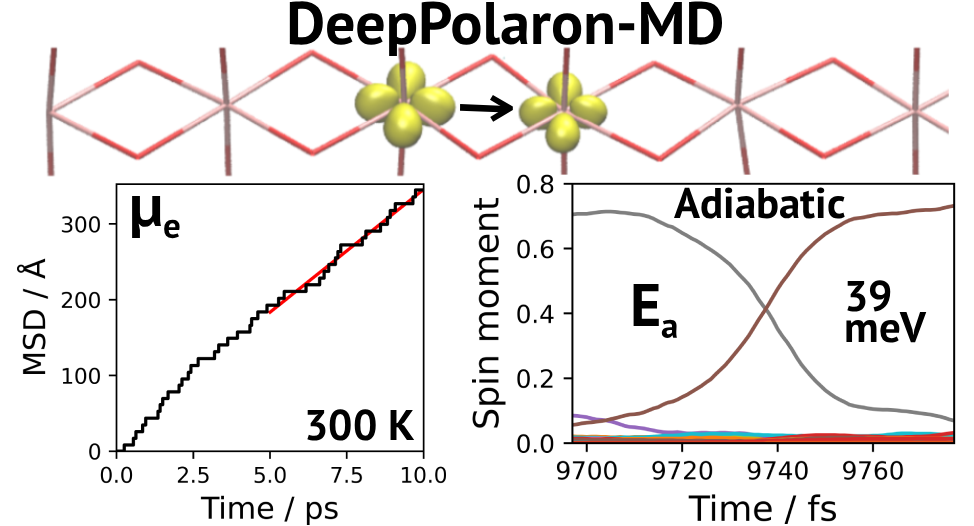}
\end{tocentry}

\clearpage
\begin{abstract}
Transition metal oxides have attracted much attention as photo(electrochemical)-catalysts but practical applications are typically hampered by their low and anisotropic charge mobility. A deep understanding of excess charge carrier transport in these materials requires a dynamical treatment of nuclear motion that goes well beyond standard approaches. Here we introduce DeepPolaron, a  machine learning framework boosting the accessible time scale of first principles molecular dynamics of adiabatic polaron transport by three orders of magnitude at a virtually negligible loss in accuracy. We apply our method to excess electron and hole transport in titanium dioxide rutile and anatase. We find that the excess electron in rutile relaxes to a polaron predominantly localized on a single Ti atom with hopping occurring only along the [001] direction, associated with an activation energy of 39 meV and a room temperature mobility of \num{4.4E-02} cm$^2$/Vs in good agreement with experiment. In contrast the hole polaron in anatase is localized on a single O atom, and due to poor O 2p orbital overlap with first nearest neighbors charge transport occurs primarily to second nearest neighbors, with a large activation energy of 139 meV resulting in a small room temperature mobility of \num{1.4E-03} cm$^2$/Vs. This work provides a finite temperature first-principles characterization of small polaron transport in rutile and anatase, with a methodology that is directly transferable to other small polaron forming materials and interfacial charge-transfer processes.
\end{abstract}

\section{Introduction} \label{Introduction}

With ever increasing global demands for energy and challenges in energy storage, the direct conversion of solar radiation into a chemical store of energy is highly desirable. Titanium dioxide (\ce{TiO2}) is an abundant, stable, and non-toxic semiconductor that has long served as a benchmark material for photoelectrochemical water splitting and related photocatalytic applications\cite{fujishima_Electrochemical_1972, schneider_Understanding_2014, herrmann_Heterogeneous_1999, ni_review_2007, fiorenza_Efficient_2019, fujishima_Titanium_2005}. However, problems remain including a large band gap leading to absorption only in UV, a low and anisotropic carrier mobility as well as short carrier lifetimes due to electron-hole recombination\cite{nakata_TiO2_2012, ohtani_Titania_2013}. While \ce{TiO2} is one of the most well studied semiconductors, a complete atomistic level understanding of charge transport, particularly under finite-temperature conditions, remains incomplete.

The two most common polymorphs of \ce{TiO2}, rutile and anatase, are understood to form small electron and hole polarons respectively. Charge transport occurs through thermally activated polaron hopping, however the intrinsic carrier mobilities remain experimentally poorly constrained, spanning three orders of magnitude from \num{1e-2} to 10 cm$^2$/Vs for the rutile electron polaron\cite{hendry_Electron_2004, yagi_Electronic_1996, tamaki_Femtosecond_2009, austin_Polarons_1969, enright_Spectroscopic_1996, bak_Defect_2003, bak_Mobility_2008}, necessitating insight from first-principles modeling. However despite extensive effort, substantial disagreement persists regarding both the magnitude of the charge-carrier mobility and the associated activation energy for polaron hopping. For rutile, electron paramagnetic resonance (EPR) measurements report an activation energy of 24 ± 5 meV for the thermal release of trapped electrons\cite{yang_Intrinsic_2013}, which has been widely interpreted in the polaron transport literature as the activation energy for polaron hopping.\cite{dai_Identification_2024, yin_Excess_2018, strand_Intrinsic_2018, morita_Models_2023, spreafico_nature_2014, elmaslmane_FirstPrinciples_2018, franchini_Polarons_2021}. In contrast, computational estimates span more than an order of magnitude, ranging from 13 to 152 meV\cite{deskins_Electron_2007, behara_Electron_2019, morita_Models_2023, birschitzky_Machine_2025, dai_Identification_2024, spreafico_nature_2014, wang_Constrained_2021}. Moreover, finite temperature effects are often neglected, and the room temperature stability of the electron polaron remains unclear\cite{elmaslmane_FirstPrinciples_2018}.

The seminal work of Deskins and Dupuis\cite{deskins_Electron_2007} used the methodology of Farazdel et al.\cite{farazdel_Electric_1990}, a conceptual precursor to constrained DFT (CDFT), with a Hubbard $U$ parameter of 10 eV, chosen to reproduce the experimental band gap, in combination with cluster models extracted from small periodic supercells to establish that the electron polaron in rutile localizes on a single Ti atom with highly anisotropic mobility: an activation energy of 90 meV for adiabatic first nearest neighbor hopping in the [001] direction, and an activation energy of 310 meV for non-adiabatic second nearest neighbor hopping in the [111] direction. More recent work by Behara and Dupuis\cite{behara_Electron_2019} with large periodic supercells further demonstrated that the activation energy varies strongly with the choice of $U$, ranging from 39 meV ($U=6$ eV) to 152 meV ($U=10$ eV), highlighting the sensitivity of electron transfer parameters to computational details.

Deskins and Dupuis\cite{deskins_Intrinsic_2009} also studied the hole polaron in anatase, finding that the hole localizes on a single O atom with adiabatic hopping to first nearest neighbors associated with an activation energy of 170 meV. However, they also found a large electronic coupling to second nearest neighbors of 480 meV which they concluded may lead to a multi-site delocalised polaronic structure. 

Optimally tuned range-separated hybrid functionals, where the fraction of exact Hartree-Fock exchange (HFX) is tuned to minimize the DFT self-interaction error (SIE), have become the standard approach to treat small polarons in recent years\cite{mckenna_Crossover_2012, blumberger_Constrained_2013, elmaslmane_FirstPrinciples_2018, ahart_Polaronic_2020, ahart_Electron_2022, palermo_Migration_2024}. Elmaslmane et al.\cite{elmaslmane_FirstPrinciples_2018} performed calculations using an optimally tuned functional PBE0 with 10.5\% HFX and large supercells, finding that the rutile electron polaron is predominantly localized on a single Ti atom with a small self-trapping energy of 25 meV, indicating that the polaron may not be stable at room temperature. For the hole polaron in anatase the same study reported a larger self-trapping energy of 210 meV, from which Carey et al.\cite{carey_Hole_2021} calculated an activation energy of 200 meV for adiabatic first nearest neighbor hopping and 120 meV for adiabatic second nearest neighbor hopping. 

Using self-interaction-corrected LDA, Dai et al.\cite{dai_Identification_2024} computed a self-trapping energy of 111 meV for the rutile electron polaron and identified it as a 3-site Ti polaron analogous to the trimeron in magnetite (\ce{Fe3O4})\cite{senn_Charge_2012}. Treating this 3-site polaron as an effective 1-site polaron for transport calculations, they obtained an activation energy of 13 meV for first nearest neighbor hopping.

To date, almost all calculations of electron transfer parameters in \ce{TiO2} have been performed using transition states obtained from DFT geometry optimization. Consequently, the thermal effect of phonons are neglected and the room temperature dynamics are approximated from static calculations. To better understand polaron dynamics under experimental operating conditions, it is necessary to develop methodologies to allow for finite temperature calculations. Quantitative analysis of the Landau–Zener transmission coefficients (Tables S4 and S9) demonstrates that the dominant hopping pathways for both the rutile electron polaron and the anatase hole polaron are firmly in the adiabatic regime, justifying the use of Born–Oppenheimer DFT-MD for modeling polaron transport in these systems.

Using PBE with a Hubbard $U$ parameter of 3.9 eV Birschitzky et al.\cite{birschitzky_Machine_2025} reported DFT-MD of the electron polaron in rutile. By training a neural network potential (NNP) and performing NNP-MD, they calculated a mobility of approximately 1.5 cm$^2$/Vs with an activation energy of 79 meV. While their work represents a major conceptual improvement over traditional static approaches, their mobility is substantially larger than the theoretical upper limit for small polaron mobility of 1 hop per lattice vibration\cite{emin_2013_polarons}. For an optical phonon frequency of  \num{2.42e13} s$^{-1}$\cite{porto_Raman_1967}, we estimate an upper limit of \num{4.1e-1} cm$^2$/Vs. In addition, their activation energy represents a somewhat surprising combination of different transport processes including first, second, third nearest neighbor hopping as well as so-called `delocalization-driven long distance hopping'. 

In this work we introduce DeepPolaron, a machine learning framework boosting the accessible time scale of first principles MD simulation of adiabatic polaron transport by three orders of magnitude at a virtually negligible loss in accuracy. By contrast to the single neural network approach of Birschitzky et al.\cite{birschitzky_Machine_2025} (termed LEOPOLD), DeepPolaron employs two coupled but independently trained neural networks, one for the potential energy surface and one for the charge population. DeepPolaron has a lower force error (10.95 meV/Å) than LEOPOLD (16.48 meV/Å) on the published rutile TiO$_2$+e$^{-}$ dataset\cite{birschitzky_Machine_2025}, providing a reliable description of adiabatic polaron dynamics.

Trained on the optimally tuned hybrid functional HSE, DeepPolaron-MD predicts that rutile electron polaron transport occurs through first nearest neighbor hopping in the [001] direction, associated with an activation energy of 39 meV and a room temperature mobility of \num{4.4e-2} cm$^2$/Vs. For the anatase hole polaron, as a consequence of poor O 2p orbital overlap to first nearest neighbors, charge transport primarily occurs through hopping to second nearest neighbors with a large activation energy of 139 meV. Carrier mobilities are determined directly from the mean-squared displacement of the charge population, and activation energies are obtained from an Arrhenius analysis, thereby avoiding some of the approximations and dependence on experimental parameters such as phonon mode frequencies that are common in prior work\cite{deskins_Electron_2007, behara_Electron_2019, morita_Models_2023, dai_Identification_2024, spreafico_nature_2014, wang_Constrained_2021}.

The remainder of the paper is organized as follows. Section \ref{Computational_methods} presents the computational methods. Sections \ref{rutile_0k} presents results for rutile, Section \ref{anatase_0k} presents the results for anatase, Section \ref{Discussion} provides a discussion of the results and concluding remarks are made in Section \ref{Conclusion}.

\section{Computational methods}  \label{Computational_methods}

To access longer time scales than are feasible with DFT-MD, and to perform MD across a range of different temperatures, we use neural network potentials (NNP). Calculation of the polaron mobility from MD trajectories requires access to the time-dependent charge population, which is not available in standard NNP architectures. We therefore introduce DeepPolaron, a machine learning framework implemented within the DeePMD-kit package.\cite{zeng_DeePMDkit_2025}. While conceptually related to the LEOPOLD framework of Birschitzky et al.,\cite{birschitzky_Machine_2025, birschitzky_Machine_2022} DeepPolaron employs two coupled neural networks trained independently on the potential energy surface and the charge population, rather than a single network. The charge population is calculated from Hirshfeld partitioning of the electron density, and the loss function incorporates terms depending on the atomic charge population, total charge, and spin, thereby enforcing charge and spin conservation during training. By contrast, LEOPOLD was trained on the charge population calculated from Ti d-state Hubbard $U$ projectors, which does not conserve total charge or spin in the published datasets.\cite{birschitzky_Machine_2025}. In addition we perform a detection of trivial crossings, instantaneous long-range polaron relocations, during DFT-MD and exclude them from DeepPolaron training data. Further implementation details are provided in Section 1 of the Supporting Information (SI).

Training data for the DeepPolaron NNPs and reference DFT-MD trajectories for polaron dynamics are obtained using the range-separated hybrid functional HSE06\cite{krukau_Influence_2006}, with the fraction of HFX modified to 22\% for rutile and 19\% for anatase. These values simultaneously reproduce the experimental band gaps and satisfy the generalized Koopmans condition with a non-linearity of less than 0.05 eV\cite{elmaslmane_FirstPrinciples_2018, ahart_Polaronic_2020}. Additional details are provided in SI Sections 2.1 and 3.1.

All calculations used the DZVP-MOLOPT-SR-GTH basis set,\cite{goedecker_Separable_1996} with explicit treatment of Ti(3s, 3p, 3d, 4s) and O(2s, 2p) valence electrons. To decrease the large computational cost of hybrid DFT calculations we use the Auxiliary Density Matrix Method (ADMM),\cite{guidon_Auxiliary_2010} with the same auxiliary basis set as Elmaslmane et al.\cite{elmaslmane_FirstPrinciples_2018}: FIT9 for Ti and cpFIT3 for O. 

Cell optimization was performed for both rutile and anatase, resulting in lattice parameters of $a=b=4.59$~\AA\ and $c=2.96$~\AA\ for rutile, and $a=b=3.77$~\AA\ and $c=9.68$~\AA\ for anatase, in good agreement with experimental values\cite{burdett_Structuralelectronic_1987}. As CP2K does not support hybrid functionals with k-points, cell optimization was performed for a $3\times3\times6$ supercell of rutile (324 atoms) and a $4\times4\times2$ supercell of anatase (384 atoms). Unless specified otherwise, these supercell sizes are used for all calculations in this work.

The polaron self-trapping energy, also known as the formation energy,
\begin{equation}
E_{\mathrm{trap}} = E (N \pm 1, \mathbf{R}_{\text{N ± 1}}) - E (N \pm 1, \mathbf{R}_{\text{N}}),
\label{eq:trapping_energy}
\end{equation}
is defined as the energy difference for the charged system with N ± 1 electrons between the geometry-optimized polaron $ E (N \pm 1, \mathbf{R}_{\text{N ± 1}})$ (where the charge is localized) and the geometry-optimized neutral ground state $ E (N \pm 1, \mathbf{R}_{\text{N}})$ (where the charge is delocalised).

To investigate finite-temperature polaron dynamics, DFT-MD was performed at 300 K for rutile and 600 K for anatase with a timestep of 1.0 fs and a Nosé-Hoover thermostat\cite{nose_unified_1984, martyna_Nose_1992}. The anatase hole polaron has a much smaller mobility than the rutile electron polaron, and therefore a higher temperature is required to provide a similar sampling of polaron hopping. Due to the substantial computational cost of HSE-MD, each system was first equilibrated for 10 ps using PBE-MD on the neutral cell. For rutile, this was followed by 20 ps of electron polaron PBE+$U$-MD ($U$=3 eV on Ti), then 10 ps of HSE(25\%)-MD, and finally a 10 ps production run with HSE(22\%). For anatase, the corresponding sequence was 20 ps of hole polaron PBE+$U$-MD ($U$=5 eV on O), 10 ps of HSE(22\%)-MD, and a 20 ps production run with HSE(19\%). No polaron hopping was observed during the PBE+$U$ and intermediate HSE MD, which is found to be beneficial for polaron equilibration prior to the production runs. While this equilibration protocol is somewhat pragmatic, it is essential to minimize the computational cost of achieving a near-equilibrium starting structure for the production HSE-MD runs. Full equilibration and statistically converged MSD calculations are performed exclusively at the DeepPolaron-MD level.

DeepPolaron NNPs were trained on structures manually selected from DFT-MD and re-computed as single point calculations. We used the PyTorch backend of the DeePMD-kit package\cite{zeng_DeePMDkit_2025} with an Adam optimizer\cite{kingma_Adam_2017}, an exponential learning rate decaying from \num{1e-3} to \num{3e-5} over 5000 steps and a total of 400000 training steps. The invariant descriptor DPA3\cite{zhang_Graph_2025} was used with a cutoff radius of 6 Å. An example training input file is provided in SI Section 1.

\begin{figure}[t!]
\includegraphics[width=1\columnwidth]{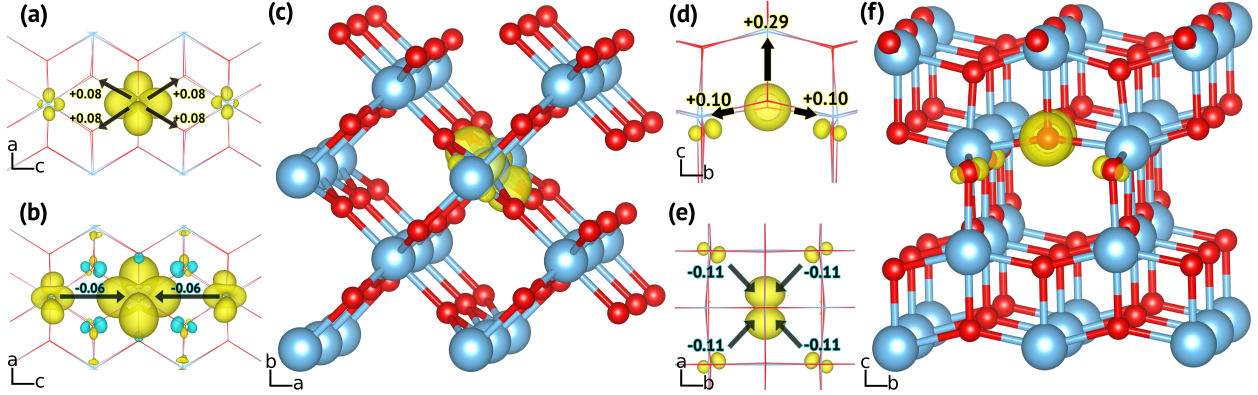}
     \caption{Polaron structure in rutile and anatase. Excess spin density for charged ground state (a-c) rutile electron polaron and (d-f) anatase hole polaron. Bond length differences with respect to neutral are shown in Angstrom. Isosurface values are (a, c-f) 0.01 e/Bohr$^{3}$ and (b) 0.003 e/Bohr$^{3}$.}
     \label{fig:polaron_structure}
\end{figure}

\begin{figure}[t!]
\includegraphics[width=1\columnwidth]{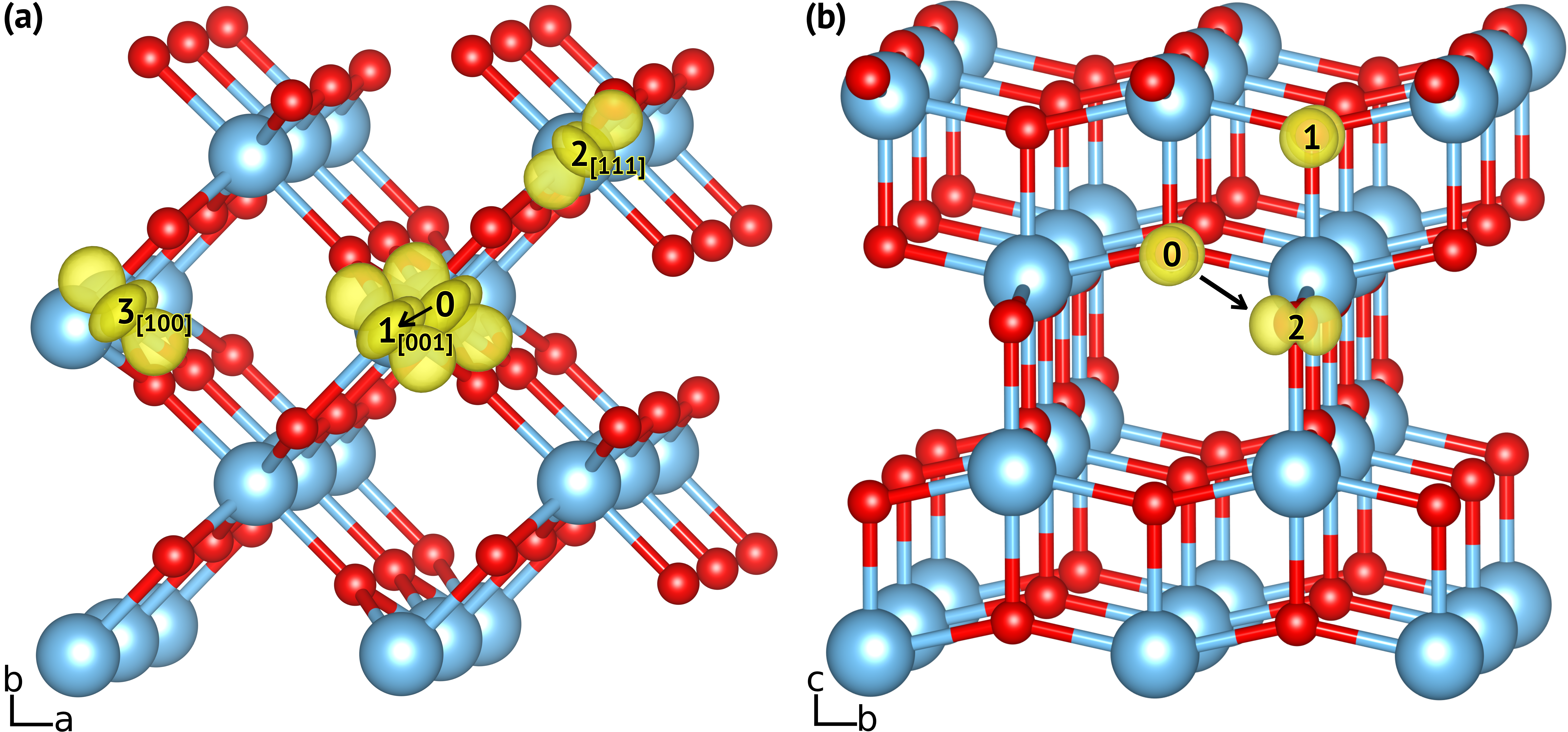}
     \caption{Polaron hopping in rutile and anatase. Excess spin density for charged ground state (a) rutile electron polaron and (b) anatase hole polaron, superimposed for the first three unique nearest neighbors for rutile and first two unique nearest neighbors for anatase. The primary hopping pathway is indicated with a black arrow: (a) first nearest neighbor hopping and (b) second nearest neighbor hopping. }
     \label{fig:polaron_hopping}
\end{figure}

\section{Results} \label{Results}

\subsection{Rutile electron polaron}  \label{rutile_0k}

The rutile electron polaron, shown in Figure \ref{fig:polaron_structure}, localizes with 68\% of the excess spin density on a single Ti atom. The polaron is stabilized by an expansion of the local Ti-O bonds, on average +0.08 \AA \ for the 4 equatorial Ti-O bond lengths in the [001] direction and an average of +0.02 \AA \ for the 2 axial Ti-O bond lengths in the [010] direction. An additional 8\% of the excess spin density is localized over each of the two first nearest neighbors in the [001] direction, which move towards the Ti atom where the polaron is localized with a decrease in the Ti-Ti distance of 0.06 \AA. The resulting polaron self-trapping energy (eq \ref{eq:trapping_energy}) is small, 96 meV. 

\begin{figure}[t!]
\includegraphics[width=1\columnwidth]{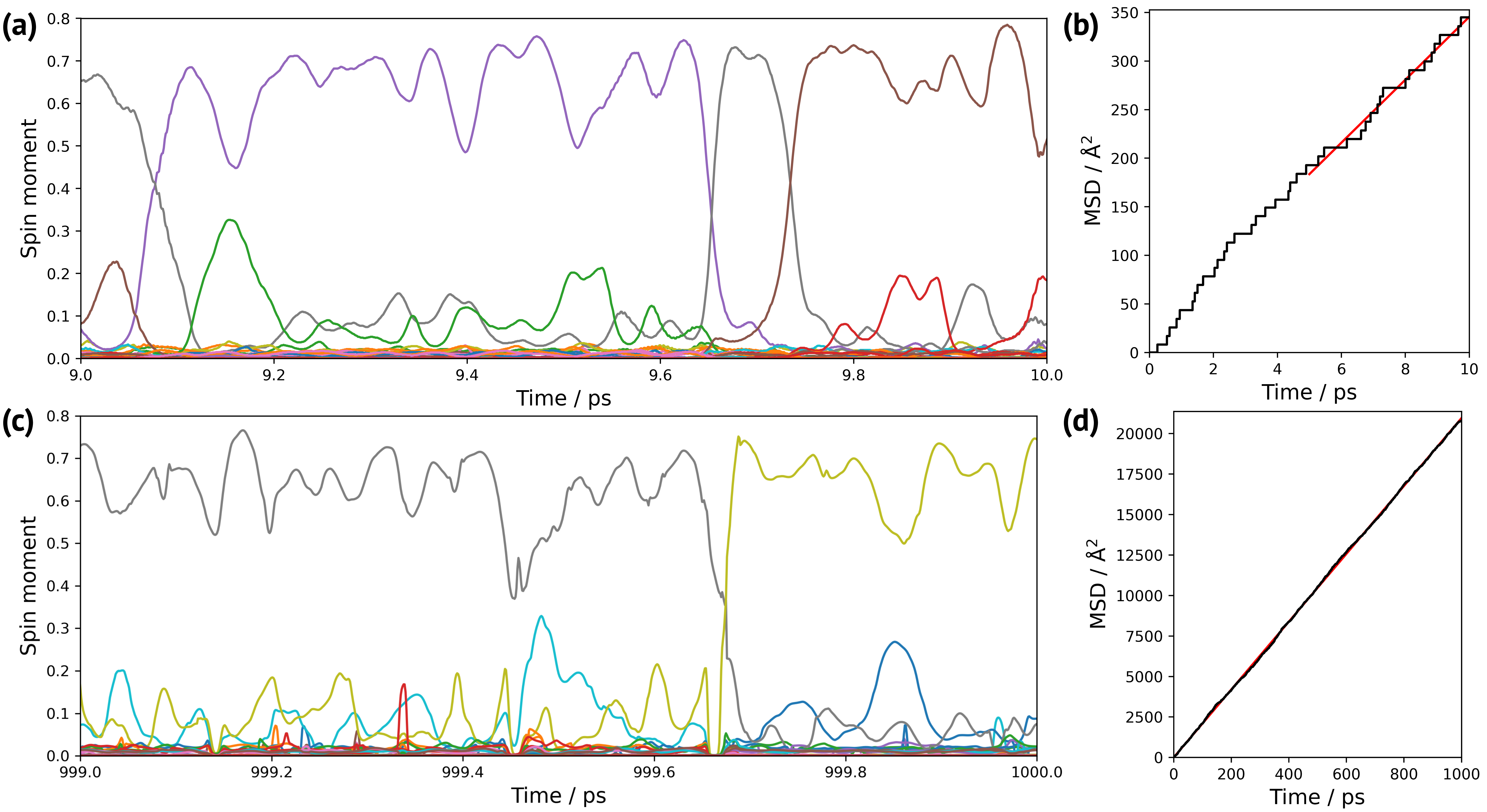}
     \caption{(a, b) DFT-MD and (c, d) DeepPolaron-MD of the electron polaron in rutile. (a, c) Spin moments color coded for each atom for the final 1 ps MD. The full 10 ps DFT-MD is shown in Figure S7. (b, d) Mean-squared displacement (MSD) of the charge population, with a diffusion coefficient calculated from a straight line fitted to the (b) last 5 ps and (d) full 1000 ps shown as a red line. }
     \label{fig:rutile_md}
\end{figure}

\begin{figure}[t!]
\includegraphics[width=1\columnwidth]{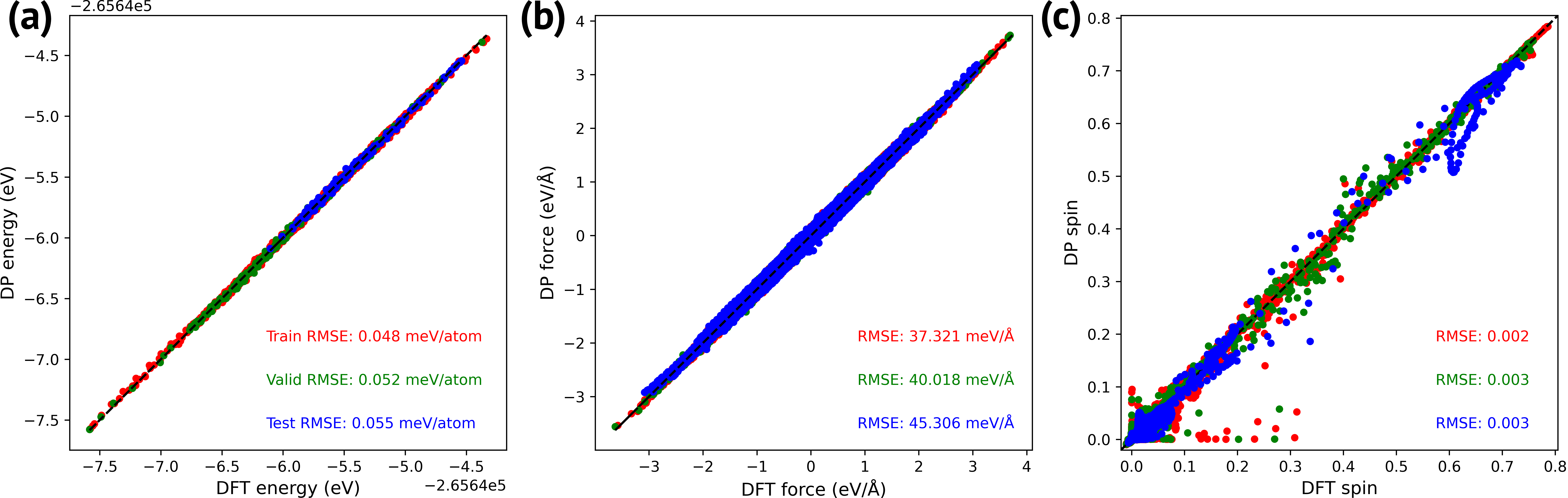}
     \caption{Training of DeepPolaron for the electron polaron in rutile. Parity plots showing a comparison of DeepPolaron (DP) and DFT (a) total energy, (b) atomic forces and (c) atomic spin moments. Red markers represent training data, green markers validation data and blue markers test data. RMSE values are shown in the bottom right of each figure, and tabulated in Table S6.}
     \label{fig:rutile_ml}
\end{figure}

The possible hopping pathways of the polaron are shown in Figure \ref{fig:polaron_hopping}, with Figure \ref{fig:rutile_md} summarizing 10 ps of DFT-MD at 300 K. The polaron localizes on a single Ti atom with an average spin moment of $0.63 \pm 0.09$, consistent with the geometry optimized value of 0.68. Thermal motion breaks the symmetry between the two first nearest neighbors, instead of equal spin moments of 0.08 they fluctuate with averages of $0.14 \pm 0.07$ and $0.05 \pm 0.02$. The polaron mid-gap state calculated from DFT-MD is $0.75 \pm 0.21$ eV below the conduction band, in good agreement with experiment\cite{austin_Polarons_1969, stoneham_Trapping_2007, cronemeyer_Infrared_1959, bogomolov_Optical_1968, setvin_Direct_2014}, and is slightly smaller than the geometry optimized value of 0.81 eV. This reduction is consistent with occasional sampling of transition state structures where the polaron is delocalised over both the initial and final sites.

Charge transport occurs exclusively through first nearest neighbor hops in the [001] direction. At the hopping transition state, the average spin moment is $0.36 \pm 0.05$, in good agreement with the value of 0.37 obtained from linear interpolation between geometry optimized polaron structures. These results indicate that thermal motion produces an ensemble of transition state structures with average properties consistent with static calculations. 

We find polaron hops are not uniformly distributed in time. Of the 39 hops, around 22 occur within correlated flurries of 2 to 3 consecutive hops. Figure \ref{fig:rutile_md} shows an example, with polaron hops occurring at 9.655 ps and 9.737 ps. The time between these hops is only 82 fs, less than two optical phonon mode periods (42 fs)\cite{porto_Raman_1967}. 

From the mean-squared displacement (MSD) of the charge population in the final 5 ps DFT-MD we estimate a mobility of \num{6.3e-2} cm$^2$/Vs. Using a pre-exponential factor in the rate equation $\nu_{\mathrm{n}} = \num{2.42e13}$~s$^{-1}$\cite{porto_Raman_1967}, this corresponds to an activation free energy of 49 meV. However, the choice of pre-exponential factor introduces significant uncertainty, and therefore the activation free energy is more reliably obtained from an Arrhenius analysis. 

Performing DFT-MD across a range of temperatures would be prohibitively expensive with HSE(22\%), therefore we trained NNPs using the DeepPolaron framework. With training data composed of 1470 frames extracted from the HSE(22\%) polaron MD, we achieve an average energy root mean square error (RMSE) of 0.06 meV/atom, a force RMSE of 45.31 meV/\AA \ and an atomic spin RMSE of 0.003 (Figure \ref{fig:rutile_ml}). Performing 1 ns DeepPolaron-MD, summarized in Figure \ref{fig:rutile_md}, we find 2409 polaron hops from which we calculate a room temperature mobility of \num{4.4e-2} cm$^2$/Vs, in good agreement with the DFT-MD estimate of \num{6.3e-2} cm$^2$/Vs given the limited sampling. 

Carrying out DeepPolaron-MD simulations at different temperatures between 150 K to 300 K we obtain the temperature dependent mobility. From an Arrhenius analysis we obtain an activation energy of 39 meV and a pre-exponential factor of \num{1.13e13} s$^{-1}$, which is smaller than the optical phonon frequency \num{2.42e13} s$^{-1}$\cite{porto_Raman_1967}. This highlights the sensitivity of electron transfer parameters to experimental values, and the motivation for the use of DeepPolaron-MD to avoid this dependency.

CDFT calculations on the rutile electron polaron (Table S3) show that polaron hopping to second and third nearest neighbors has high activation barriers, and are unlikely to be observed on the ps time scale in agreement with results from present DFT-MD and DeepPolaron-MD.

\subsection{Anatase hole polaron}  \label{anatase_0k}

\begin{figure}[t!]
\includegraphics[width=1\columnwidth]{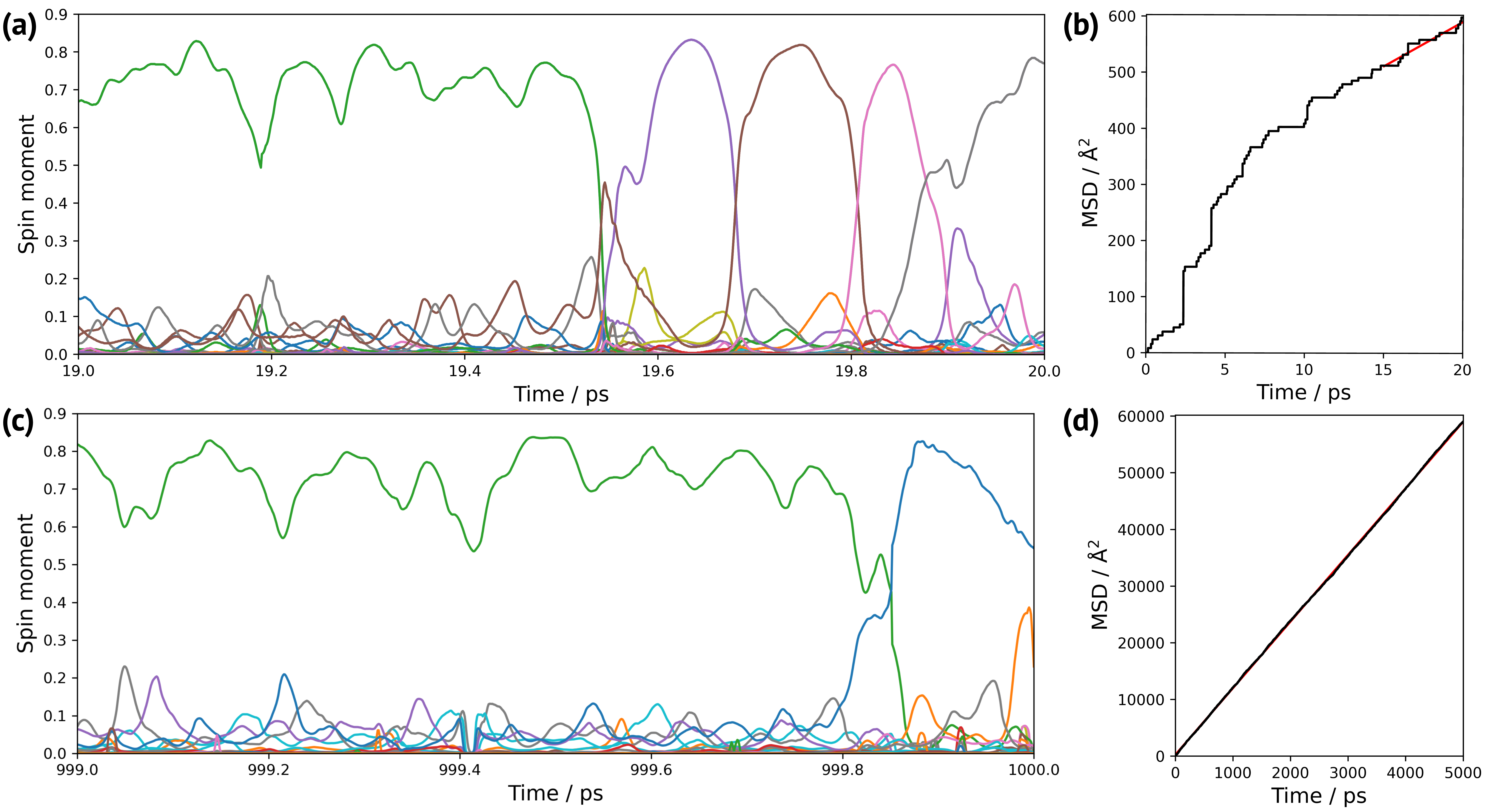}
     \caption{(a, b) DFT-MD and (c, d) DeepPolaron-MD of the hole polaron in anatase. (a, c) Spin moments color coded for each atom for the final 1 ps MD. The full 20 ps DFT-MD trajectory is shown in Figure S25. (b, d) Mean-squared displacement (MSD) of the charge population, with a diffusion coefficient calculated from a straight line fitted to the (b) last 5 ps and (d) full 5000 ps shown as a red line.}
     \label{fig:anatase_md}
\end{figure} 

\begin{figure}[t!]
\includegraphics[width=1\columnwidth]{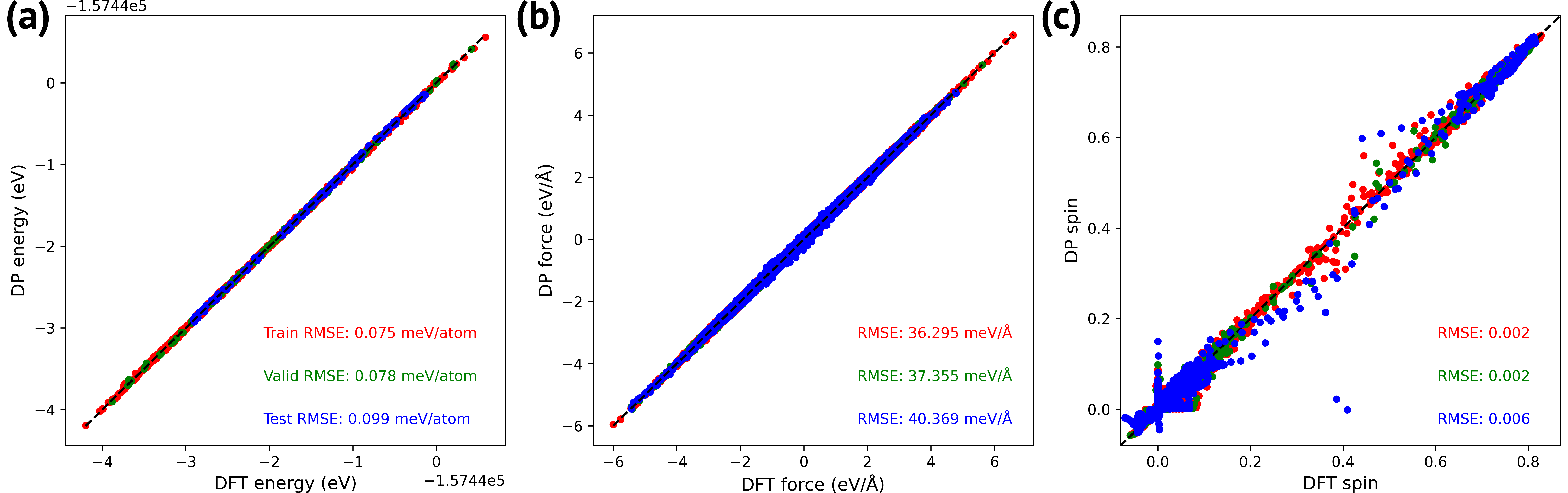}
     \caption{Training of DeepPolaron for the hole polaron in anatase. Parity plots showing a comparison of DeepPolaron (DP) and DFT (a) total energy, (b) atomic forces and (c) atomic spin moments. Red markers represent training data, green markers validation data and blue markers test data. RMSE values are shown in the bottom right of each figure, and tabulated in Table S13.} 
     \label{fig:anatase_ml}
\end{figure}

The anatase hole polaron, shown in Figure \ref{fig:polaron_structure}, localizes with 78\% of excess spin density on a single O atom. An additional 5\% of the excess spin density is distributed over each of the four second nearest neighbors, which are displaced toward the polaron site with a reduction in O–O distance of 0.11 \AA. In contrast to the typical contraction of the local coordination shell in response to the formation of a hole polaron\cite{ahart_Polaronic_2020}, the local O-Ti bonds elongate to maximize the O 2p orbital overlap with second nearest neighbors: +0.29 \AA \ for the axial O-Ti and +0.10 \AA \ for the two equatorial O-Ti. The resulting self-trapping energy (eq \ref{eq:trapping_energy}) is large, 426 meV. 

The possible hopping pathways of the polaron are shown in Figure \ref{fig:polaron_hopping}, with Figure \ref{fig:anatase_md} summarizing 20 ps of DFT-MD at 600 K. The polaron localizes on a single O atom with an average spin moment of $0.70 \pm 0.09$, consistent with the geometry optimized value of 0.78. The polaron mid-gap state calculated from DFT-MD is $1.23 \pm 0.53$ eV above the valence band, smaller than the geometry optimized value of 1.66 eV, consistent with occasional sampling of transition state structures where the polaron is delocalised over both the initial and final state.

In 20 ps DFT-MD we observe 2 first nearest neighbor hops, 52 second nearest neighbor hops and 5 instantaneous hops over large distances between 5 and 10 \AA. Similar to rutile, we find polaron hops are not uniformly distributed in time. Of the 59 hops, around 39 occur within flurries of 2 to 5 consecutive hops. Figure \ref{fig:anatase_md} shows an example, where 4 polaron hops occur between 19.540 ps and 19.879 ps with consecutive hops separated by less than 140 fs. This corresponds to fewer than 4 optical phonon mode periods (38 fs).\cite{gonzalez_Infrared_1997}

The 5 long-range instantaneous hops introduce non-linearity in the MSD of the charge population, complicating the calculation of the mobility. Extending the DFT-MD to 20 ps (compared to 10 ps for rutile), a linear regime is identified in the final 5 ps, from which we estimate a mobility of \num{1.5e-2} cm$^2$/Vs. Using the pre-exponential factor $\nu_{\mathrm{n}} = \num{2.66e13}$~s$^{-1},$\cite{gonzalez_Infrared_1997} the corresponding activation free energy is 133 meV. 

Training DeepPolaron on 1170 frames extracted from HSE(19\%)-MD including only second nearest neighbor hopping, we achieve an average energy RMSE of 0.10 meV/atom, a force RMSE of 40.37 meV/\AA \ and an atomic spin RMSE of 0.006 (Figure \ref{fig:anatase_ml}). Performing 5 ns DeepPolaron-MD, summarized in Figure \ref{fig:anatase_md}, we find 7607 polaron hops from which we calculate a mobility of \num{9.5e-3} cm$^2$/Vs, in good agreement with the mobility estimated from DFT-MD of \num{1.5e-2} cm$^2$/Vs given the limited sampling. Carrying out DeepPolaron-MD simulations at different temperatures between 300 K to 600 K we obtain the temperature dependent mobility. From an Arrhenius analysis we obtain an activation energy of 139 meV, and a similar pre-exponential factor of \num{2.32e13} s$^{-1}$ to the optical phonon frequency for anatase \num{2.66e13} s$^{-1}$\cite{gonzalez_Infrared_1997}.

CDFT calculations on the anatase hole polaron (Table S9) show that the electronic coupling to second nearest neighbors is substantially larger than to first nearest neighbors, leading to a larger mobility, in agreement with results from present DFT-MD and DeepPolaron-MD. 

\clearpage

\begin{figure}[t!]
\includegraphics[width=0.9\columnwidth]{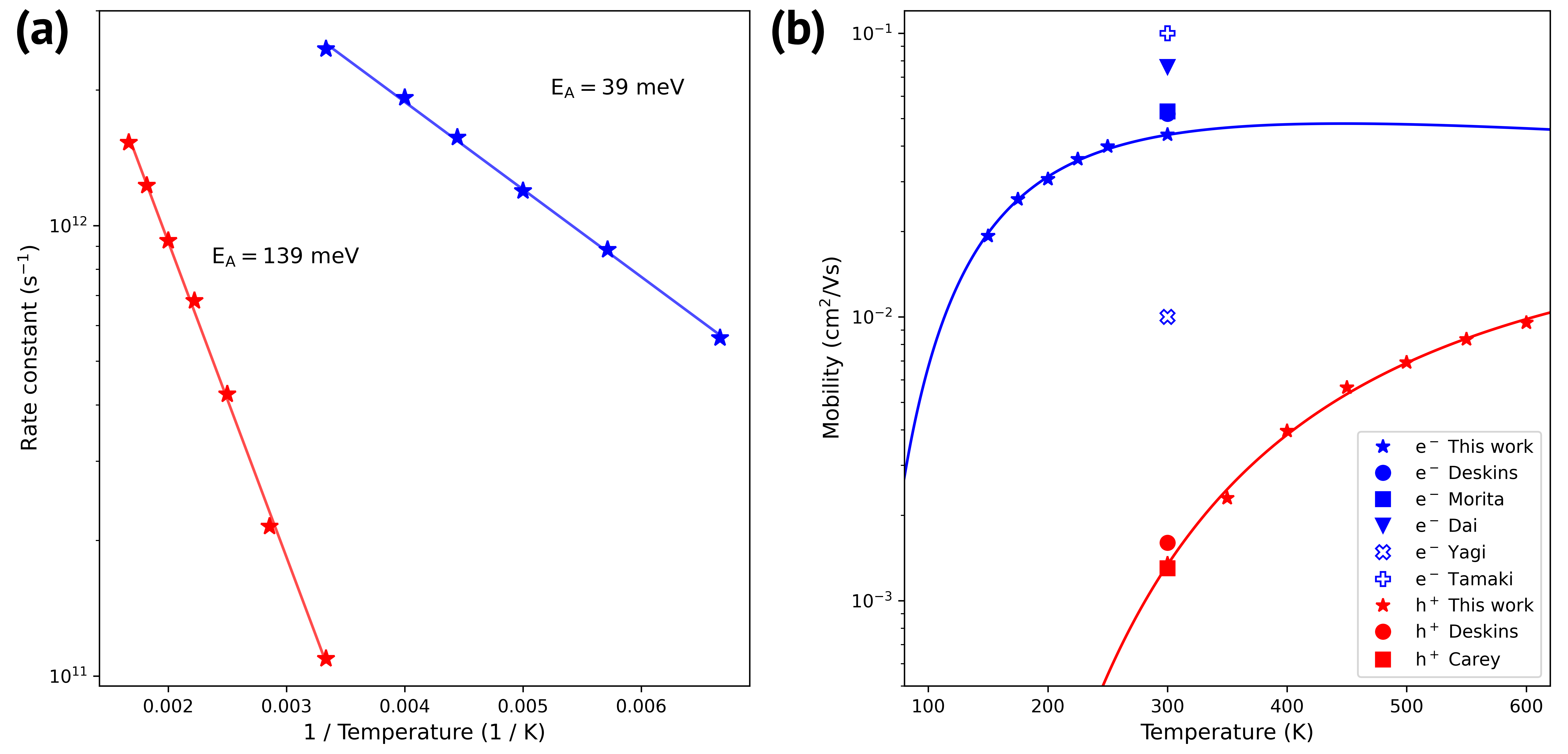}
     \caption{Rate constant and mobility as a function of temperature. (a) Rate constants calculated from DeepPolaron-MD fitted with the Arrhenius equation. (b) Mobilities calculated from DeepPolaron-MD fitted with the Einstein equation. Calculated mobilities are indicated with solid filled markers, experimental mobilities as unfilled markers. Only literature mobilities compatible with small polaron hopping are included. }
     \label{fig:arrhenius_mobility}
\end{figure}

\begin{table*}[t!]
\footnotesize
\begin{threeparttable}
\caption{Summary of results and comparison with literature. To facilitate direct comparison, both the activation energy ($\Delta A^{\ddagger}_{\text{lit}}$) and mobility cited in the paper ($\mu_{\text{lit}}$) are re-calculated in the adiabatic limit of large electronic coupling\cite{spencer_Confronting_2016} assuming a transmission coefficient of 1.0. The site multiplicity used to estimate the analytic diffusion coefficient is 1\cite{ahart_Electron_2022}. Literature results are presented in chronological ordering by method.}
\centering
\begin{tabular*}{\textwidth}{@{\extracolsep{\fill}}lllllll}
  \hline
   \multicolumn{7}{c}{\textbf{Rutile electron polaron}}   \\
   \hline
  \hline
Source & $H_{\mathrm{{ab}}}$   & $\lambda$    &  $\Delta A^{\ddagger}_{\text{lit}}$   & $\Delta A^{\ddagger}$ & $\mu_{\text{lit}}$  & $\mu$     \\
 & (meV)   &  (meV)   &   (meV)   & (meV) & (cm$^2$/Vs)  &  (cm$^2$/Vs)    \\
 \hline
This work &    &   & & 39 &  & \num{4.4E-02} \\
Linear DFT \cite{deskins_Electron_2007} &   200 &  1152 & 90 & 123 &  \num{5.2E-02}  & \num{1.7E-03} \\
NEB DFT \cite{morita_Models_2023} &  21  &  230 & 58 & 56 &  \num{5.3E-02} & \num{2.2E-02} \\
Linear DFT \cite{dai_Identification_2024} &    &   & 13 &  &   \num{5.1E-1}  & \num{1.2E-1}  \\
NNP-MD\cite{birschitzky_Machine_2025} &    &   & 79 &  &   $\approx$ \num{1.5} & \\
EPR\cite{yang_Intrinsic_2013} &    &   & 24 ± 5 &  &  &  \\
Hall\cite{austin_Polarons_1969} &    &   &  &&  1 &   \\
Hall\cite{yagi_Electronic_1996} &    &   &  & &  \num{1E-02}   & \\
Spectroscopy\cite{hendry_Electron_2004} &    &   &  & &  1 &   \\
Spectroscopy\cite{tamaki_Femtosecond_2009} &    &   &  && \num{1E-01}  &   \\
\hline
\multicolumn{7}{c}{\textbf{Anatase hole polaron}}   \\
   \hline
  \hline
This work  &     &   &  & 139 & & \num{1.4e-3} \\
Linear DFT\cite{deskins_Intrinsic_2009} &  480  &  2040 & & 143 &   & \num{1.6e-3} \\
Linear DFT\cite{carey_Hole_2021} &  113  &  930 & 120 & 133 & $\approx$ \num{1e-3} & \num{2.4e-3} \\
 \hline
\end{tabular*}
\label{tab:summary_of_results}
  \end{threeparttable}
\end{table*}

\clearpage

\section{Discussion} \label{Discussion}

Table \ref{tab:summary_of_results} compares our results with values reported in the literature, with Figure \ref{fig:arrhenius_mobility} presenting rate constants and mobilities as a function of temperature. A direct comparison of different calculations is difficult due to the different methodologies used, therefore we alleviate this somewhat by comparing to activation energies and mobilities re-calculated in the adiabatic limit of large electronic coupling\cite{spencer_Confronting_2016}.

DeepPolaron-MD predicts first nearest neighbor hopping of the electron polaron in rutile, associated with an activation energy of 39 meV and a room temperature mobility of \num{4.4E-02} cm$^2$/Vs, a mechanism consistent with all previous static calculations of electron transfer parameters\cite{deskins_Electron_2007, morita_Models_2023, dai_Identification_2024}. In contrast, recent work from Birschitzky et al.\cite{birschitzky_Machine_2025} using PBE+$U$-MD and NNP-MD identified a combination of 74.3\% first nearest neighbor hopping, 0.2\% second nearest neighbor hopping, 25.5\% third nearest neighbor hopping and so-called `delocalization-driven long distance hopping'. The resulting mobility of approximately 1.5 cm$^2$/Vs is almost two orders of magnitude larger than current and previous work\cite{deskins_Electron_2007, morita_Models_2023, dai_Identification_2024}, and substantially exceeds the theoretical upper bound for small polaron mobility of \num{4.1e-1} cm$^2$/Vs estimated from the optical phonon frequency. While the cause for the `delocalization-driven long range hopping' is unknown, we speculate that they could be a manifestation of trivial crossing events, a problem that is well known in the non-adiabatic dynamics simulation of polaron transport in materials\cite{giannini_Flickering_2020}. Because DFT-MD is performed within the Born-Oppenheimer approximation on the adiabatic ground state, the total potential energy is minimized at each timestep without continuous propagation of the electronic wavefunction. Consequently, thermal fluctuations may spuriously drive the polaron across large distances when another site becomes momentarily lower in energy. While we do not observe trivial crossings for the rutile electron polaron, we do for the anatase hole polaron and find that excluding these crossings from the DeepPolaron training data is essential for obtaining a linear MSD and a well-defined mobility. 

Despite substantial methodological differences, our room-temperature mobility of \num{4.4e-2}~cm$^2$/Vs from DeepPolaron-MD agrees closely with that reported by Deskins and Dupuis (\num{5.2e-2}~cm$^2$/Vs) using the methodology of Farazdel et al.\cite{farazdel_Electric_1990} with cluster models extracted from small periodic supercells\cite{deskins_Electron_2007}. However, despite this similar mobility, our activation energy of 39 meV is substantially smaller than their value of 90 meV. This reflects compensating approximations in their estimation of the mobility from the calculated electron transfer parameters. First, the pre-exponential factor in the Arrhenius rate expression: Deskins and Dupuis\cite{deskins_Electron_2007} use the optical phonon frequency for rutile (\num{2.42e13} s$^{-1}$\cite{porto_Raman_1967}), while our Arrhenius fit gives a smaller value of \num{1.13e13} s$^{-1}$. Second, the analytic estimate of the diffusion coefficient: we compute $D$ directly from the mean squared displacement of the charge population, whereas analytic treatments approximate $D = r^2ik/2$ where r is the hopping distance, i is the site multiplicity and k is the rate constant\cite{rosso_initio_2003, iordanova_Charge_2005, ahart_Electron_2022}. Deskins and Dupuis\cite{deskins_Electron_2007} used a site multiplicity of 2 to account for the two nearest neighbors along [001] and omit the factor $1/2$, leading to a four-fold overestimate of $D$ on this term alone. Together with the prefactor difference, this gives a factor of more than 8 in the mobility. Re-calculating their activation energy in the adiabatic limit of large electronic coupling\cite{spencer_Confronting_2016} also results in an increase from 90 meV to 123 meV, further decreasing the mobility. Applying all three corrections to the data of Deskins and Dupuis\cite{deskins_Electron_2007} gives a recomputed mobility of \num{1.7e-3}~cm$^2$/Vs, more than an order of magnitude smaller than their reported value.

Morita et al.\cite{morita_Models_2023} also reported a comparable mobility of \num{5.3e-2}~cm$^2$/Vs using HSE(25\%), however their semi-quantitative calculation of electron transfer parameters yields a small electronic coupling of 29 meV at the linearly interpolated transition state, decreasing further to 21 meV upon refinement using nudged elastic band (NEB). While Morita et al.\cite{morita_Models_2023} do not perform analysis to confirm whether the polaron hopping is adiabatic or non-adiabatic, we use their reported electron transfer parameters to compute Landau–Zener transmission coefficients\cite{wu_Extracting_2006} of 0.8 and 0.7, indicating non-adiabatic polaron hopping. This contrasts sharply with our results and those of Deskins and Dupuis\cite{deskins_Electron_2007}, which consistently demonstrate adiabatic hopping as a result of the substantial Ti 3d orbital overlap along [001]. To provide a direct comparison, we apply the methodology of Morita et al.\cite{morita_Models_2023} to our own calculations (Table S4), and find the same trend: an electronic coupling of 97 meV at the linearly interpolated transition state for first nearest neighbor hopping that decreases to 12 meV upon refinement using NEB, with the transmission coefficient correspondingly decreasing from 1.0 to 0.3. We therefore find that this semi-quantitative calculation of electron transfer parameters does not reliably reproduce the electronic coupling, and that the agreement between the mobility of Morita et al.\cite{morita_Models_2023} and our work and that of Deskins and Dupuis\cite{deskins_Electron_2007} is coincidental.

Elmaslmane et al.\cite{elmaslmane_FirstPrinciples_2018} investigated the electron polaron in rutile using optimally tuned PBE0(11.5\%), finding a self-trapping energy of only 25 meV, smaller than $k_{\mathrm{B}}T = 26$ meV at 300 K, suggesting that the polaron may be unstable at room temperature. To test this hypothesis we performed PBE0(11.5\%)-MD (Figure S11), finding a highly unstable polaron with a mobility of \num{1.3} cm$^2$/Vs, exceeding the theoretical upper bound for small polaron mobility of \num{4.1e-1} cm$^2$/Vs. This highlights an important limitation of optimal tuning, that the non-linearity is not sufficiently sensitive to the HFX fraction. The non-linearity of PBE0 is less than 50 meV between 11.5\% and 14\% HFX, while the self-trapping energy varies between 25 meV and 101 meV. In contrast the band gap varies between 2.82 eV and 3.04 eV, with an experimental value of 3.02 eV\cite{amtout_Optical_1995}. We therefore recommend using the experimental band gap as an additional constraint, as performed here with HSE(22\%) (Figure S2). 

Similar to rutile, for anatase we find that the mobility is anisotropic, with only second nearest neighbor hopping contributing significantly to the mobility as a result of poor O 2p orbital overlap with first nearest neighbors. In contrast, Deskins and Dupuis\cite{deskins_Intrinsic_2009} concluded that there is adiabatic hopping only to first nearest neighbors, with second nearest neighbors forming a multi-site delocalised polaronic structure. Re-calculating their activation energies in the adiabatic limit of large electronic coupling\cite{spencer_Confronting_2016} we obtain values of 207 meV and 143 meV for first and second nearest neighbor hopping, the latter which is the primary transport mechanism with a mobility of \num{1.6e-3} cm$^2$/Vs. With this improved calculation of the activation energy both our mobility (\num{1.4e-3}~cm$^2$/Vs) and activation energy (139 meV) are in very close agreement with that of Deskins and Dupuis\cite{deskins_Intrinsic_2009}. 

Carey et al.\cite{carey_Hole_2021} also performed calculations for the anatase hole polaron, using a similar semi-quantitative methodology to that of Morita et al\cite{morita_Models_2023}. For transition states obtained from linear interpolation, they report adiabatic first and second nearest neighbor hopping activation energies of 200 meV and 120 meV, yielding a mobility on the order of \num{1e-3} cm$^2$/Vs. Using their reported electron transfer parameters we compute transmission coefficients of 1.0 for both pathways, confirming adiabatic hopping. Their results are therefore in good agreement with the transport mechanism and electron transfer parameters obtained in our work and that re-computed from Deskins and Dupuis\cite{deskins_Intrinsic_2009}.

Standard small polaron hopping theories assume that the lattice relaxes fully between successive hops, so that each hop is independent of the previous one\cite{EMIN1969439}. In contrast, we find that many polaron hops occur within correlated flurries. This is consistent with the work of Emin, who identified that for adiabatic polaron hopping the lattice may not have time to fully relax between hopping events, and that until relaxation is complete further hops are energetically facilitated\cite{emin_Correlated_1970, emin_Lattice_1971}. The flurries found for the anatase hole polaron are much longer, 2 to 5 consecutive hops with a mean intra-flurry hop gap of 110 fs, than for the rutile electron polaron, 2 to 3 consecutive hops with a mean intra-flurry hop gap of 86 fs. This is consistent with the larger activation energy of the anatase hole polaron (139 meV compared to 39 meV) that is associated with a longer lattice relaxation time, and is only somewhat compensated for by the higher simulation temperature of 600 K.

Direct comparison with experiment is complicated by the presence of defects and uncertainties in estimating the carrier density. Nevertheless, our calculated mobility for the electron polaron in rutile, \num{4.4e-2} cm$^2$/Vs, lies within the broad range of reported experimental values \num{1e-2} to 10 cm$^2$/Vs\cite{hendry_Electron_2004, yagi_Electronic_1996, tamaki_Femtosecond_2009, austin_Polarons_1969, enright_Spectroscopic_1996, bak_Defect_2003, bak_Mobility_2008}, and our activation energy of 39 meV is in good agreement with EPR measurements of $24\pm5$ meV\cite{yang_Intrinsic_2013}. To the best of our knowledge, there are no experimental measurements of mobility or activation energy for the hole polaron in anatase at room temperature. Further experimental work, for example direct mobility measurements on p-type anatase samples, is required to verify the predicted small polaron transport mechanism and the quantitative mobility calculated in this work.

\section{Conclusion} \label{Conclusion}

We have presented a comprehensive study of small polaron transport in the two most common polymorphs of titanium dioxide, rutile and anatase. Unlike previous studies, which often perform semi-quantitative calculations of electron transfer parameters or neglect thermal effects, we have developed a modern, state-of-the-art machine learning approach trained on optimally tuned hybrid functionals to obtain a fully quantitative description of polaron transport. Through calculation of the mobility directly from the mean squared displacement of the charge population, and performing an Arrhenius analysis to calculate the activation energy, we remove some of the approximations and dependence on experimental parameters such as the phonon mode frequency common in previous static electron transfer theory calculations.

Our results show that the electron polaron in rutile localizes predominantly on a single Ti atom, with hopping along the [001] direction associated an activation energy of 39 meV and a room-temperature mobility of \num{4.4E-02} cm$^2$/Vs. In contrast, the hole polaron in anatase localizes on a single O atom, and due to poor O 2p orbital overlap with first nearest neighbors charge transport occurs primarily to second nearest neighbors with a large activation energy of 139 meV leading to a small room temperature mobility of \num{1.4E-03} cm$^2$/Vs. In contrast to the conventional model of independent polaron hopping, we find that the many polaron hops occur through correlated flurries. 

We find that spurious long-range hopping can occur during DFT-MD as a consequence of trivial surface crossings, and we hypothesize that this might have caused an overestimate of the rutile electron polaron mobility in recent work\cite{birschitzky_Machine_2025}. By training a neural network without including long-range hops in the training data, we have shown for the anatase hole polaron that it is possible to calculate a well defined mobility consistent with static electron transfer theory calculations.  

We have demonstrated that neural network potentials can accurately reproduce polaron dynamics in \ce{TiO2}, enabling the study of finite-temperature transport on previously inaccessible timescales. This methodology is broadly applicable to other small polaron forming materials and interfacial charge-transfer processes in the adiabatic transport regime, offering a powerful tool for the emerging field of \textit{ab initio} electrochemistry.

\begin{acknowledgement}

Via our membership of the UK’s HEC Materials Chemistry Consortium, which is funded by EPSRC (EP/L000202, EP/R029431), this work used the ARCHER UK National Supercomputing Service (http://www.archer.ac.uk). Additional computational resources were provided by the Westlake HPC Center.

\end{acknowledgement}

\begin{suppinfo}

The Supporting Information is available free of charge

\begin{itemize}
  \item Details of DeepPolaron, optimal tuning, finite size effects, electron transfer parameters from CDFT and semi-quantitative methods, details of DFT-MD and DeepPolaron-MD, direct comparison between DeepPolaron and LEOPOLD, comparison between HSE and PBE0 (PDF)
\end{itemize}

\end{suppinfo}

\bibliography{zotero.bib} 

\end{document}